# New Discoveries in Stars and Stellar Evolution through Advances in Laboratory Astrophysics

Submitted by the

American Astronomical Society Working Group on Laboratory Astrophysics
http://www.aas.org/labastro

Nancy Brickhouse - Harvard-Smithsonian Center for Astrophysics
nbrickhouse@cfa.harvard.edu, 617-495-7438

John Cowan - University of Oklahoma
cowan@nhn.ou.edu, 405-325-3961

Paul Drake* - University of Michigan
rpdrake@umich.edu, 734-763-4072

Steven Federman - University of Toledo
steven.federman@utoledo.edu, 419-530-2652

Gary Ferland - University of Kentucky
gary@pa.uky.edu, 859-257-879

Adam Frank - University of Rochester
afrank@pas.rochester.edu, 585-275-1717

Eric Herbst - Ohio State University
herbst@mps.ohio-state.edu, 614-292-6951

Keith Olive - University of Minnesota
olive@physics.umn.edu, 612-624-7375

Farid Salama - NASA/Ames Research Center
Farid.Salama@nasa.gov, 650-604-3384

Daniel Wolf Savin - Columbia University
savin@astro.columbia.edu, 1-212-854-4124,

Lucy Ziurys - University of Arizona
lziurys@as.arizona.edu, 520-621-6525

*Editor

# 1. Introduction

As the Stars and Stellar Evolution (SSE) panel is fully aware, the next decade will see major advances in our understanding of these areas of research. To quote from their charge, these advances will occur in studies of "the Sun as a star, stellar astrophysics, the structure and evolution of single and multiple stars, compact objects, SNe, gamma-ray bursts, solar neutrinos, and extreme physics on stellar scales."

Central to the progress in these areas are the corresponding advances in laboratory astrophysics, required to fully realize the SSE scientific opportunities within the decade 2010-2020. Laboratory astrophysics comprises both theoretical and experimental studies of the underlying physics that produces the observed astrophysical processes. The 6 areas of laboratory astrophysics, which we have identified as relevant to the CFP panel, are atomic, molecular, solid matter, plasma, nuclear physics, and particle physics.

In this white paper, we describe in Section 2 the scientific context and some of the new scientific opportunities and compelling scientific themes which will be enabled by advances in laboratory astrophysics. In Section 3, we discuss some of the experimental and theoretical advances in laboratory astrophysics required to realize the SSE scientific opportunities of the next decade. As requested in the Call for White Papers, Section 4 presents four central questions and one area with unusual discovery potential. Lastly, we give a short postlude in Section 5.

# 2. Scientific context, opportunities, and compelling themes

Stellar structure has been of interest since at least the early 20$^{th}$ century work of Eddington and his contemporaries. It took a bit longer to begin to understand stellar evolution and the role of stars in the production of the elements. This, however, was only the start. With the advent of observatories that can detect energetic radiation from x-rays to gamma rays to neutrinos to millimeter radiation revealing molecular lines, our awareness of the scope of phenomena in stars, PNe and SNe has broadened vastly. The discovery of gamma ray bursts and other dramatic and unexpected events has stimulated great continuing interest. Meanwhile, simulations have made it possible to integrate our apparent understanding of stellar dynamics and have turned out to show how little we understand the details of stars and their behavior. Laboratory astrophysics can contribute to this knowledge in myriad ways, ranging from the measurement of specific cross sections, reaction rates, and spectral features needed to interpret data and model stars, to the examination of physical processes that cannot be approximated well in the simulations.

The quest to understand in quantitative detail the origin of the elements and Galactic chemical evolution remains an enduring theme. Multiple uncertainties exist regarding both the long-term structure and evolution of stars and the brief but essential explosion phase. We know far more about the interior of the sun than of any other star, thanks to helioseismology. One of the things we know at present is that the predicted location of the boundary of the solar convective zone disagrees with that observed by 13 standard deviations (Basu and Antia, 2008). We know that neutrinos provide the energy that powers core-collapse SNe, but how they do so remains mysterious. We know that SNe are asymmetric and that they develop complex internal structure, but not in detail how they do so. We know that stars typically are magnetized, but do not understand in detail the role of magnetic fields in their evolution.

A second theme is the role of stars, PNe and SNe as immediate drivers of nearby structure. As the star evolves, so does the stellar wind, introducing structured material around the star, for example producing the varied morphologies of PNe. The SNe is clearly not uniform and spherical, and its nonuniformities carry structure into the surrounding medium. In the longer term, both the shock and the structured material from the SNe interact with the surrounding medium. Then the structure we observe has contributions from both the star and the ISM.

A third theme is the exploration of the extreme physics present in SSE. The environment around various compact objects has magnetic fields strong enough to alter atomic structure, electron-positron pair plasmas, relativistic shock waves and jets, and photoionized plasmas. The emissions from such systems are often strongly altered from those that would be produced in less extreme environments. Interpreting them requires models of complex, extreme physics. Knowing the models are correct requires experiments in relevant regimes.

A fourth theme is the structure of stellar atmospheres above the photosphere. That stellar coronae are ~ 1000 times hotter than the underlying photospheres has long been known but not understood. A major clue is the observation in the sun that the abundances of coronal elements with a low first ionization potential are enhanced over their photospheric values. Such abundance fractionization is observed in many stars, but some stars exhibit no such effect or the reverse effect. Detailed models attribute these effects to ponderomotive interactions with Alfven waves (Laming 2009).

### 3. Required advances in Laboratory Astrophysics

#### 3.1. Atomic Physics

*Line identification* from emission or absorption spectra is the first step in analyzing and modeling their origins. This requires accurate and complete wavelengths across the electromagnetic spectrum. Understanding the properties of an observed cosmic source also depends on accurate knowledge of the underlying atomic processes. Oscillator strengths and transition probabilities are critical to a wide variety of temperature and abundance studies. Quantum calculations of collisional excitation rate coefficients, accurate to better than 10% and benchmarked with experiments (e.g., Chen et al. 2006), are needed for diagnostics of coronal and shocked plasmas. Accurate fluorescence yields for K alpha lines of iron group elements, such as Cr and Mn, may help to determine the detonation mechanism in SN explosions (Badenes et al. 2008). Many existing data for the heavier elements are still notoriously unreliable. The current limitations on the atomic data available for mid-Z elements make it difficult to determine the nature of the *r*-process.

*Modeling ionization structure* requires accurate data on many processes. Models of electron-ionized gas need high-temperature, density-dependent dielectronic recombination (DR) rates, becoming possible with $3^{rd}$ and $4^{th}$ generation light sources. Electron-impact ionization data need to be improved; existing measurements and calculations are often scant. Charge exchange (CX) interactions with H, $H^+$, He and $He^+$ are important for many systems, but few modern calculations or laboratory measurements exist at the relevant temperatures. Data are also needed for elements such as Se and Kr in order to study nucleosynthesis in the progenitors of PNe.

Understanding photoionization needs contributions from both atomic physics and plasma physics. Models need low-temperature DR rates, for which theory is challenging and laboratory measurements with light sources may provide the only reliable data. One can also produce photoionized plasmas to benchmark models by using an intense x-ray source to irradiate a volume containing relevant species, and measuring the ionization balance and other properties (Foord et al., 2006). The coming decade will see much more such work.

*Stellar opacities* contribute significantly to the questions and discrepancies that remain in understanding stellar evolution, and are an area where atomic physics and plasma physics are both essential. "Pulsed power" devices have used axial currents to create plasmas at the densities and temperatures, which are present in the solar convection zone. This has enabled direct measurement of the x-ray opacity of iron, showing some discrepancies from the values used in present-day solar models (Bailey et al., 2009). In the next decade experiments will produce and exploit plasmas corresponding to increasing depth in the sun and will explore a more complete range of elements and conditions.

*Integrated modeling* of stars and supernovae requires all the above. This has recently (Bryans et al. 2009) led to the most stringent spectroscopic constraints to date for models of abundance fractionization and coronal heating. Line emission of the various ions in SNe can be used to test explosion models (Laming and Hwang 2004) and nucleosynthesis models. The observed distribution of elements within a SNR can constrain the processes occurring during the explosion (Akiyama et al. 2002). This requires accurately taking into account line emission, non-equilibrium ionization, radiative cooling and hydrodynamics. The resulting constraints are only as good as the underlying atomic data in the models. Improved atomic data are needed in order to improve them.

### 3.2. Molecular Physics

*Identifying molecules and their abundances* absolutely requires high-resolution laboratory spectroscopy. Cool stellar atmospheres exhibit a range of molecular optical and infrared spectra. Circumstellar shells of evolved stars, both AGB and supergiants, are known to be rich in molecular compounds, as evidenced by millimeter and infrared observations. The carbon-rich envelope of the AGB star IRC+10216 is one of the richest molecular sources in the Galaxy. Even in PNe, molecular material persists. Understanding the underlying chemistry and its relationship to dust formation in the stellar envelope is crucial in evaluating mass loss from evolved stars. Studies of elemental and isotopic ratios in circumstellar molecules have also enabled observational tests of models of nucleosynthesis.

The spectroscopic study of circumstellar molecules, many which exhibit exotic structures that cannot be produced in large abundance in the laboratory, requires the development and application of state-of-the-art ultra-sensitive spectroscopic instruments. Construction and implementation of these instruments is costly and time-consuming, and data production from these devices cannot be turned on and off at will. Efficient utilization requires continued support. Detecting the presence of a species, however, is not sufficient since it must be reconciled with other physical properties of the surrounding medium. It is important to untangle the detailed chemical reactions and processes leading to the formation of new molecules in the complex environments of stellar atmospheres and ejecta. This requires the use of state-of-the-art experimental facilities. These data, together with quantum chemical calculations, will establish credible chemical models of stellar environments, thus guiding future astronomical searches of hitherto unobserved molecular species.

### 3.3. Solid Material

*Dust identification* and the understanding of dust evolution in stellar environments depend on having a systematic catalog of observable features useful for identification of specific dust components and molecular precursors. Laboratory spectroscopy of minerals and carbonaceous materials provide this information; multi-waveband studies are particularly useful for confirming identifications made from only a single feature. As an example, a feature near 21 microns observed with Spitzer in the Carina nebulaa, the SN remnant Cas A, and in some H II regions is likely associated with dust grains composed of FeO and a mix of silicates (e.g. Rho et al. 2008). The wavelength centroid varies amongst the objects observed, suggesting either coagulation or a difference in the mixture of components.

### 3.4. Plasma Physics

*Explosion hydrodynamics* is uncertain for several reasons, one of which is that calculations which fully resolve the turbulent dynamics will not be possible for several more decades. One can observe the dynamics in experiments that are well scaled to explosion conditions (Kuranz et al., 2009). Some experiments like the observations have found unanticipated outward penetration of inner material (Drake et al. 2004). The next decade will see further progress on this issue.

*Shock breakout* observations from SNe are now possible and will soon become common (e.g., Soderberg et al., 2008). The radiative shock that emerges is in a novel regime, not readily modeled even in dedicated computer codes. Experiments have produced radiative shocks in this regime, and will proceed during the next decade to benchmark computer simulations of such systems and to seek unanticipated dynamics.

*Extreme physics* is present in GRBs, which produce Lorentz factors of a few hundred and radiate strongly. Present-day relativistic lasers can produce the same Lorentz factors in systems of the same dimensionless scale (number of skin depths) as GRBs and only a very few orders of magnitude higher in density. This will enable experiments that are directly relevant to GRBs. Experiments with electron-positron pair plasmas will also become feasible in this next decade.

*Magnetic fields* in laboratory plasmas are proving to be larger and more complex than anticipated (Rygg et al., 2008), as may indeed also be true in stars. The combination of experiments and 3D modeling will produce large advances in understanding such fields over the next decade. In addition, understanding the effect of strong magnetic fields on atomic structure is essential to the analysis of spectra from the vicinity of magnetized compact objects, including neutron stars and black holes. Ggauss fields, large enough to have significant effects, are produced now (Wagner et al. 2004); the next decade will see 10 Ggauss.

*Dynamos* in stars are also poorly understood. Experiments are now testing some dynamo models in electrically conducting fluids (Monchaux et al. 2007), in turbulent plasmas (Ji et al. 1994), and in liquid metal (Spence et al. 2007). In the next decade, further progress is expected in elucidating mechanisms in generating magnetic field in MHD media and also in plasmas, and thus provide much needed understanding on the origin of magnetic field in stars and galaxies.

*Magnetic reconnection* occurs in solar flares and forming stars at rates orders of magnitude faster than classical theories can explain. Only recently have proposed faster mechanisms begun to be tested experimentally. For example, the conjectured Hall effect was verified (Ren et al., 2005). In the next decade, further progress is expected in the areas of reconnection in three dimensions, particle acceleration during reconnection, as well as reconnection rate scaling on system size.

What is required to realize the potential contributions in this area to astrophysics is for the relevant agencies to treat laboratory astrophysics as an important component of national research in plasmas, including but not limited to magnetized and high-energy-density plasmas.

### 3.5. Nuclear Physics

*Stellar Modeling and Neutrino Physics:* A decades-long effort to measure the reactions of the pp-chain and CN-cycle led to a calibrated standard solar model predicting patterns of neutrino fluxes inconsistent with experiment. This in turn motivated experiments that established that 2/3rds of the solar flux was in heavy-flavor neutrinos. More precise constraints on both the nuclear S-factors and the flavor physics will be coming from experiments. Newly precise solar neutrino experiments such as SNO+ have the potential (by measuring the CN-neutrino component of the solar flux) to determine directly the metallicity of the solar core. This is important because current solar metallicities that are inconsistent with interior solar sound speeds.

*Explosive astrophysics:* Approximately half of the elements heavier than iron were synthesized in core-collapse SN, by rapid neutron capture. One of the main goals of the planned Facility for Rare Ion Beams (FRIB) will be to produce the rare isotopes involved in the r-process for the first time, so that their masses and beta decay lifetimes can be measured. This will enable the nucleosynthetic output of SNe to be used to constrain aspects of the explosion, such as the dynamic timescale for the ejecta and the freezeout radius. These studies will also provide clues to the nature of the first stars and earliest nucleosynthesis in the Galaxy.

*Supernova physics:* Laboratory nuclear astrophysics provides critical input to SN modeling, including the Gamow-Teller strength distributions that govern neutrino-matter interactions and nucleosynthesis. The results will be incorporated into multi-D SN models now under development. By detecting and interpreting neutrinos from the next galactic SN, one may be able to constrain the unknown third mixing angle $\theta_{13}$, and to isolate a new aspect of the MSW mechanism, oscillations altered by an intense neutrino background.

*Neutron stars:* The structure of neutron stars impacts problems as diverse as the phases of QCD at high density (e.g., color-flavor locking) and the gravitational wave forms needed for the future data analysis. A key uncertainty is the precise value of the nuclear symmetry energy, which is related to the pressure of a neutron gas. A major experimental program at the Jefferson Laboratory to determine the neutron distribution of a heavy nucleus will soon address this.

*Nuclear astrophysics in heavier stars:* There are major uncertainties in our understanding of the nuclear physics governing massive star evolution. The most critical parameter in the hydrostatic evolution is the $^{12}C(\alpha,\gamma)$ cross section, which determines the C to O mass ratio, and thus influences nucleosynthesis. There are several experimental efforts underway to further constrain this cross section, which is complicated due to the contributions of subthreshold resonances.

*Transient thermonuclear reactions:* Thermonuclear explosions following accretion on neutron stars are responsible for the X-ray bursts studied by observatories such as Beppo-SAX, Chandra, XMM-Newton, RXTE, and INTEGRAL. The details of these bursts are poorly understood. The observables depend on the nuclear physics of neutron-deficient nuclei participating in the $\alpha$p- and rp processes. FRIB will make significant progress in this area, specifically by the reacceleration of beams produced by fragmentation and gas stopping.

### 3.6 Particle Physics

*New, very light particles* are predicted by certain theories in particle physics. For example, the axion appears in models addressing the so-called strong CP problem. Such axions might be produced in stars and could dramatically affect their lifetime in the red giant phase. Many key constraints on axion properties come from stellar evolution. Several laboratory experiments will search for axions by looking for the conversion to photons in strong magnetic fields.

### 4. Four central questions (●) and one area (❖) with unusual discovery potential

- What is the detailed, time-dependent structure of stars, and how does this affect their evolution and the consequences of slow neutron capture?
- What are the explosion mechanisms of SNe and what is the impact of these on nucleosynthesis?
- What are the mechanisms of mass loss from intermediate mass stars and how does such mass loss impact the structure and chemical evolution of the ISM?
- What is the actual behavior of energetic events such as GRBs and how do these impact cosmic-ray acceleration?

- ❖ The role of magnetic fields in stars and stellar evolution. Because they are difficult to measure, even outside stars, and difficult to model, which requires three dimensions for any realism, laboratory studies have the potential to make seminal discoveries.

### 5. Postlude

Laboratory astrophysics and complementary theoretical calculations are part of the foundation for our understanding of SSE and will remain so for many generations to come. From the level of scientific conception to that of the scientific return, it is our understanding of the underlying processes that allows us to address fundamental questions in these areas. In this regard, laboratory astrophysics is much like detector and instrument development; these efforts are necessary to maximize the scientific return from astronomical observations.